\documentstyle[preprint,aps]{revtex}

\begin{document}
\draft

\preprint{SLAC-PUB-6502}

\title{
THE FORM FACTORS OF THE NUCLEONS
}
\author{Felix Schlumpf}
\address{
Stanford Linear Accelerator Center\\
Stanford University, Stanford, California 94309
}
\date{May 1994}
\maketitle

\begin{abstract}
We demonstrate that a relativistic constituent quark model can give
nucleon form factors that agree well with recent, accurate measurements.
The relativistic features of the model and the specific form of
the wave function are essential for the result.
\end{abstract}


\narrowtext

The study of the electromagnetic form factors of the nucleons is
of fundamental importance in understanding the nucleon structure.
The form factors contain all the information about the deviation
from pointlike structure of the charge and magnetic current
distributions of the nucleons. Recent measurements~[1]
from Rosenbluth separations in elastic electron-proton and
quasielastic electron-deuteron scattering have doubled the $Q^2$
range of previous data on the nucleon form factors and
reduced the error bars in the region of overlap.
None of the existing models is in good agreement with all form factor
results at all values of $Q^2$, although for several models, the fit could
be improved by adjusting free parameters. In addition, in the near future
new accelerators like CEBAF will get precise data on all form factors
up to 6 GeV$^2$. Thus it seems that there is a need for reliable
theoretical predictions of the nucleon form factors for moderately large
values of $Q^2$. In the present work we investigate a relativistic
constituent quark model in the light of the recent data~[1].

We formulate the constituent quark model on the light-cone [2].
The wave function is constructed as the product of a
momentum wave function, which is
spherically symmetric and invariant under permutations,
and a spin-isospin wave function,
which is uniquely determined by SU(6)-symmetry requirements.  A
Wigner (Melosh) rotation~[3]
is applied to the spinors,
so that the wave function of the proton is an eigenfunction
of $J^2$ and $J_z$ in
its rest frame~[4].  For the momentum wave function we
choose a simple function of
the invariant mass ${\cal M}$ of the quarks:
\begin{equation}
\psi({\cal M}^2) = N(1+{\cal M}^2/\beta^2)^{-p}
\end{equation}
where $\beta$ sets the scale of the nucleon size and $p=3.5$.
 The invariant mass ${\cal M}$ can be written as
\begin{equation}
{\cal M}^2 = \sum_{i=1}^3 \frac{\vec k_{\perp i}^2+m^2}{x_i}
\end{equation}
where we used the longitudinal light-cone
momentum fractions $x_i=p_i^+/P^+$ ($P$
and $p_i$ are the nucleon and quark momenta,
respectively, with $P^+=P_0+P_z$).
The internal momentum variables $\vec k_{\perp i}$ are given by
$\vec k_{\perp
i}=\vec p_{\perp i}-x_i \vec P_\perp$ with the
constraints $\sum \vec k_{\perp i}=0$
and $\sum x_i=1$.
It has been shown [5] that observables for $Q^2=0$ are independent of the
wave function $\psi$.

The Dirac and Pauli form factors
$F_1(Q^2)$ and $F_2(Q^2)$ of the nucleons are
given by the spin-conserving and the spin-flip vector current
$J^+_V$ matrix elements ($Q^2=-q^2$)
\begin{eqnarray}
F_1(Q^2) &=& \langle p+q,\uparrow | J^+_V | p,\uparrow \rangle , \\
(Q_1-i Q_2) F_2(Q^2) &=& -2M\langle p+q,\uparrow | J^+_V | p,
\downarrow \rangle .
\end{eqnarray}
The Sachs form factors shown in the figures~1 to 5  are given by
\begin{eqnarray}
G_E(Q^2) &=& F_1(Q^2) - \frac{Q^2}{4M^2} F_2(Q^2) , \\
G_M(Q^2) &=& F_1(Q^2) +  F_2(Q^2) .
\end{eqnarray}
The measurements can be roughly described by the dipole fit $G_D(Q^2) =
(1+Q^2/0.71)^{-2}$: $G_{Ep}(Q^2) \sim G_{Mp}(Q^2)/\mu_p \sim
G_{Mn}(Q^2)/\mu_n \sim G_D(Q^2);
G_{En}(Q^2) \sim 0$.

The only parameters used in Ref.~[2] are the constituent quark mass $m$ and
the scale parameter $\beta$. For $m=263$ MeV and $\beta=607$ MeV we get
the dashed line in figures~1 to 5. The form factor $G_{Ep}$ contradicts
the new data (while it was still fine  with the old data). By including
anomalous magnetic moments for the quarks~[6] we can fix this problem. The
continuous lines in figures~1 to 5 give the form factors for the values
$m= 240$ MeV, $\beta=670$ MeV, $F_{2u}=-0.035$, and $F_{2d}=0.015$.
These are the
optimal values, and the precise measurements show some deviations for the
$G_{Mn}$ which cannot be resolved by changing parameters. Additional
physics like higher Fock states (pion cloud, gluons, strange quarks)
contribute to the form factors.

\acknowledgments
It is a pleasure to thank Fritz Coester and Stan Brodsky for
stimulating discussions.
This work was supported in part by the Schweizerischer Nationalfonds and
in part by the Department of Energy, contract DE-AC03-76SF00515.

\begin{figure}
\caption{Proton form factor $G_E$ as calculated in the relativistic
constituent quark model compared with data from Ref.~[1]. Solid
line, calculation with nonzero $F_{2q}$; dashed line,
calculation with $F_{2q}=0$.}
\end{figure}

\begin{figure}
\caption{Proton form factor $G_M$ as calculated in the relativistic
constituent quark model compared with data from Ref.~[1]. The
same line code as in Fig.~1 is used.}
\end{figure}

\begin{figure}
\caption{The ratio of the proton form factors $F_2$ and $F_1$ as
calculated in the relativistic
constituent quark model compared with data from Ref.~[1]. The
same line code as in Fig.~1 is used.}
\end{figure}

\begin{figure}
\caption{Neutron form factor $G_E$ as calculated in the relativistic
constituent quark model compared with data from Ref.~[1]. The
same line code as in Fig.~1 is used.}
\end{figure}

\begin{figure}
\caption{Neutron form factor $G_M$ as calculated in the relativistic
constituent quark model compared with data from Ref.~[1]. The
same line code as in Fig.~1 is used.}
\end{figure}

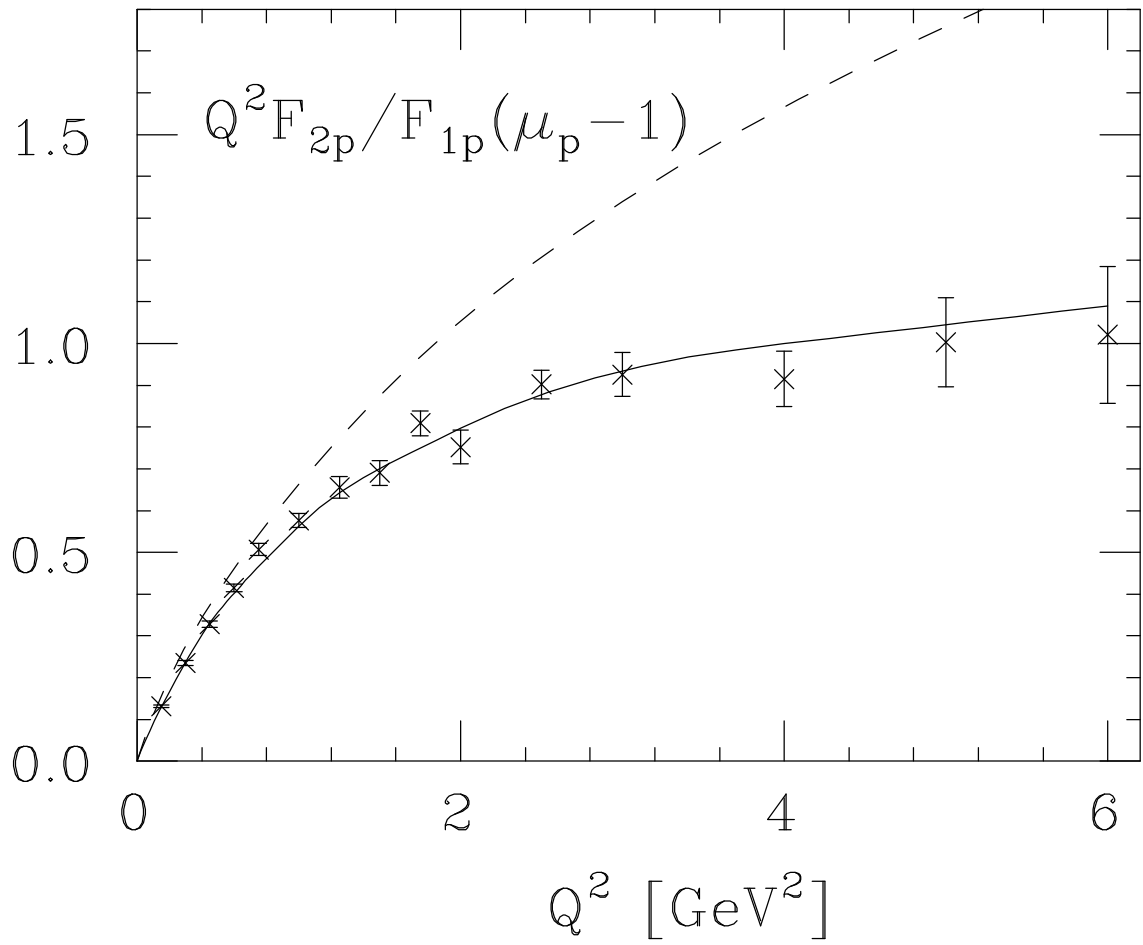
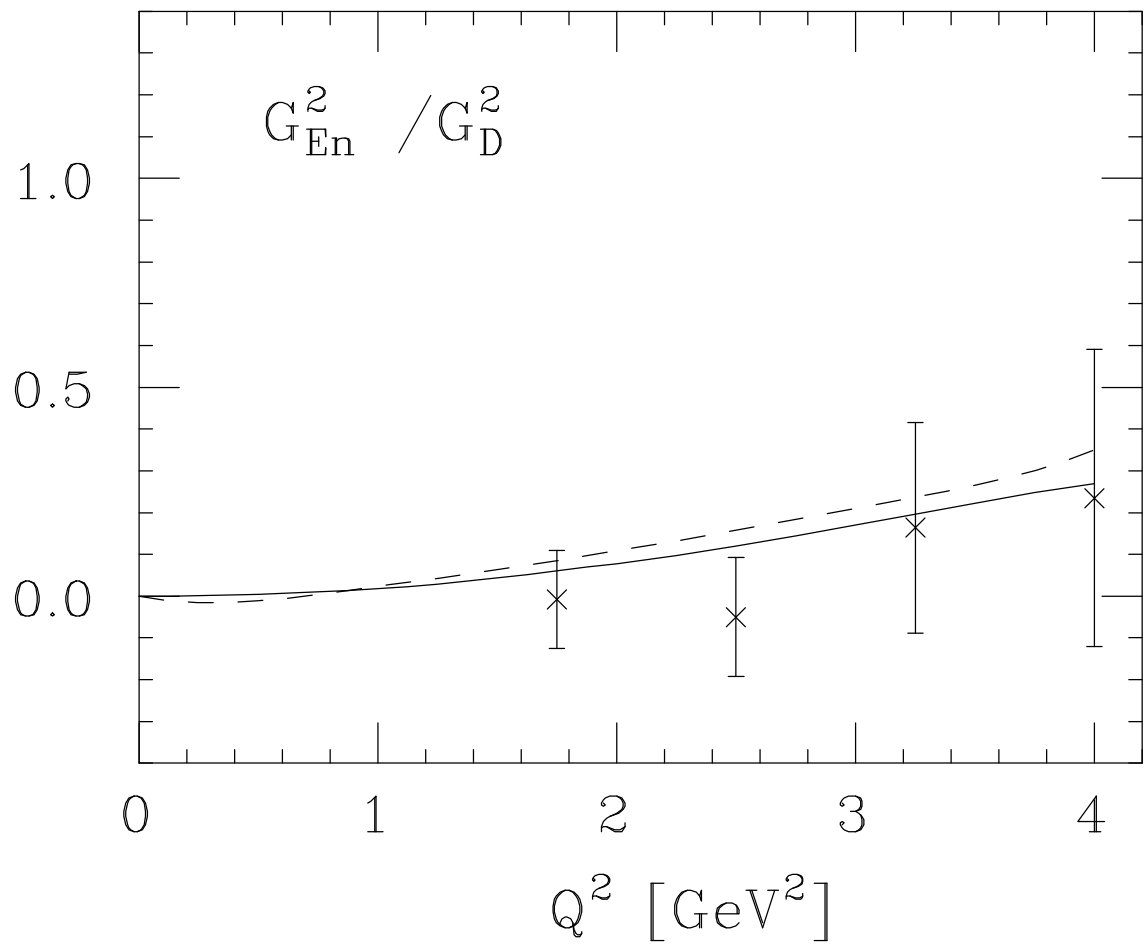
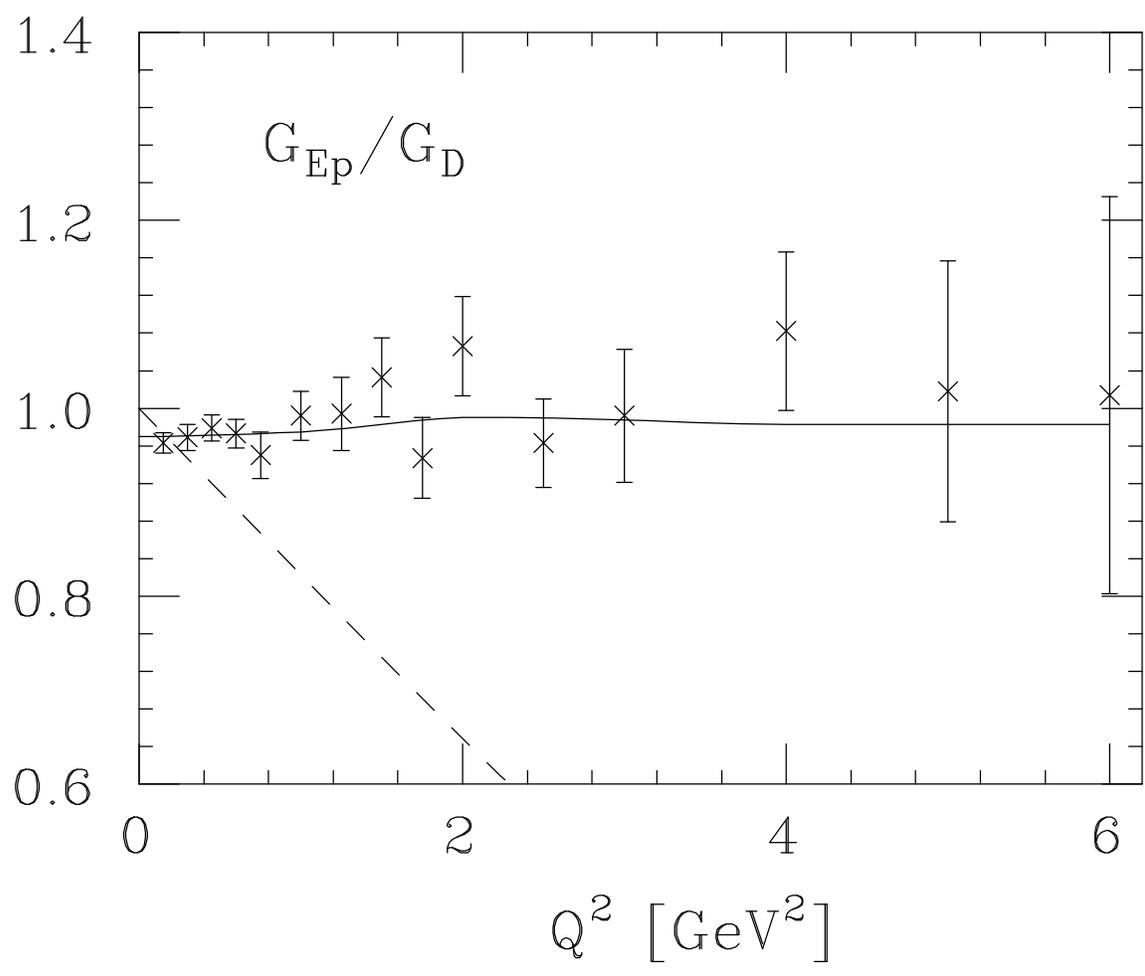
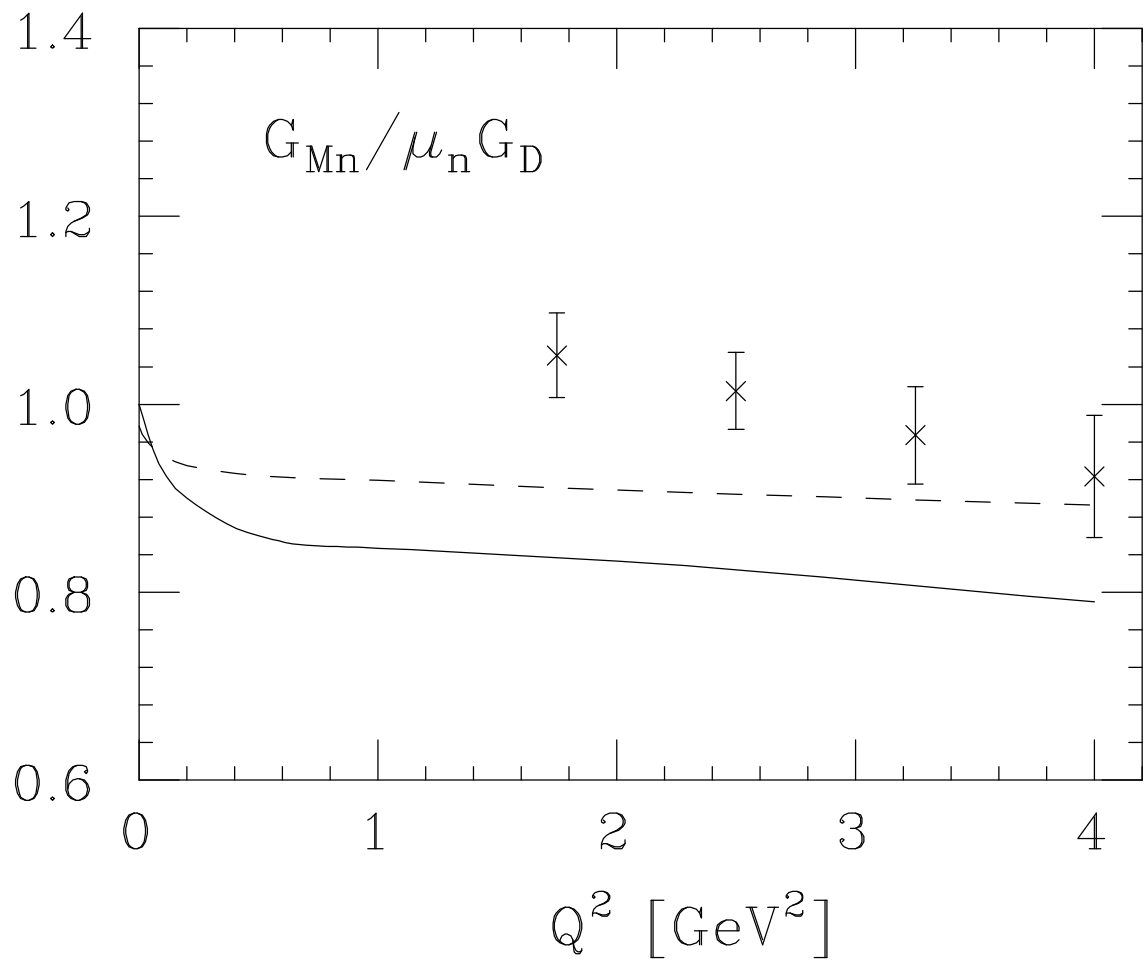
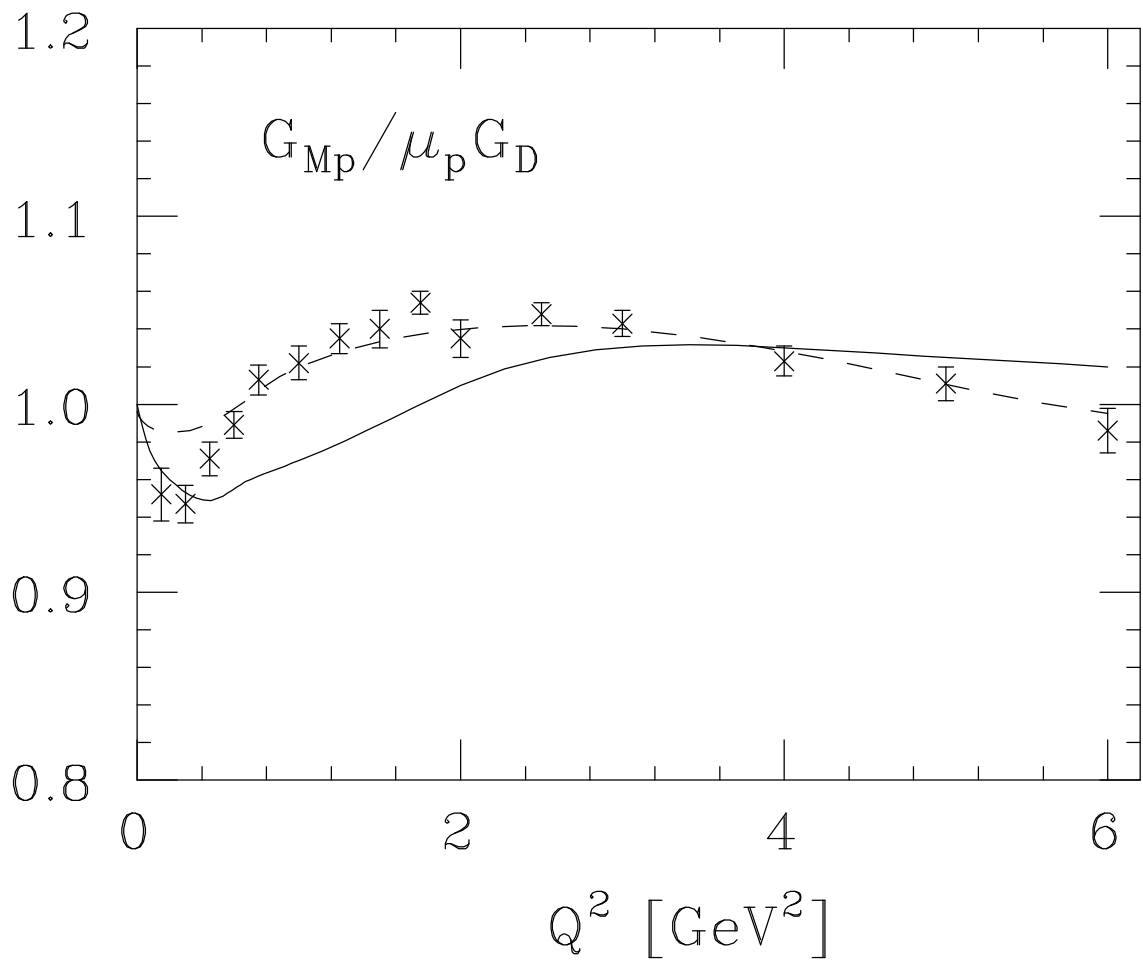
\end{document}